\documentclass[journal,10pt]{IEEEtran} 
\makeatletter
\def\endthebibliography{%
	\def\@noitemerr{\@latex@warning{Empty `thebibliography' environment}}%
	\endlist
}
\makeatother

\usepackage{cite}

\usepackage[pdftex]{graphicx}

\usepackage[caption=false,font=footnotesize]{subfig}

\usepackage{amsmath}
\usepackage{amsfonts}
\usepackage{bm}
\usepackage{filecontents}
\usepackage{amssymb}
\usepackage{color}
\usepackage{xcolor}

\usepackage{epsfig}

\usepackage{xurl}

\usepackage{algpseudocode}
\usepackage{algorithm, tabularx}

\usepackage{lettrine}

\usepackage{lipsum}

\usepackage{siunitx}

\usepackage{soul,color}

\usepackage{mathrsfs}

\usepackage{array}

\usepackage[inline]{enumitem}

\usepackage{breqn}

\usepackage{multirow}
\usepackage{array}
\newcolumntype{L}[1]{>{\raggedright\let\newline\\\arraybackslash\hspace{0pt}}m{#1}}
\newcolumntype{C}[1]{>{\centering\let\newline\\\arraybackslash\hspace{0pt}}m{#1}}
\newcolumntype{R}[1]{>{\raggedleft\let\newline\\\arraybackslash\hspace{0pt}}m{#1}}

\usepackage{amsthm}

\newlength{\maxwidth}
\newcommand{\algalign}[2]% #1 = text to left, #2 = text to right
{\makebox[\maxwidth][r]{$#1{}$}${}#2$}

\makeatletter
\newcommand{\multiline}[1]{%
	\begin{tabularx}{\dimexpr\linewidth-\ALG@thistlm}[t]{@{}X@{}}
		#1
	\end{tabularx}
}
\makeatother

\theoremstyle{remark}

\algdef{SE}[SUBALG]{Indent}{EndIndent}{}{\algorithmicend\ }%
\algtext*{Indent}
\algtext*{EndIndent}

\begin{document}
	
	\title{{Free-Space Optical Communications for\\6G Wireless Networks: {Challenges, Opportunities, and Prototype Validation}}}

	\author{Hong-Bae Jeon,~\IEEEmembership{Graduate Student Member,~IEEE}, Soo-Min Kim,~\IEEEmembership{Graduate Student Member, IEEE},\\Hyung-Joo Moon,~\IEEEmembership{Graduate Student Member, IEEE}, Do-Hoon Kwon, Joon-Woo Lee,\\Jong-Moon Chung,~\IEEEmembership{Senior Member, IEEE}, Sang-Kook Han,~\IEEEmembership{Senior Member, IEEE},\\Chan-Byoung Chae,*~\IEEEmembership{Fellow,~IEEE}, and Mohamed-Slim Alouini,~\IEEEmembership{Fellow,~IEEE}
		
	%	\thanks{This work is supported by InstituteIITP funded by the Korea government (MSIT) under Grant 2019-0-00685.}
\thanks{H.-B. Jeon, S.-M. Kim, H.-J. Moon, and C.-B. Chae* (corresponding author) are with the School of Integrated Technology, Yonsei University, Seoul 03722, Rep. of Korea (e-mail: \{hongbae08, sm.kim, moonhj, cbchae\}@yonsei.ac.kr).}% <-this % stops a space
\thanks{D.-H. Kwon, J.-W. Lee, J.-M. Chung, and S.-K. Han are with the School of Electrical and Electronic Engineering, Yonsei University, Seoul 03722, South Korea (e-mail:\{ehgns222, junu0809, jmc, skhan\}@yonsei.ac.kr).}
\thanks{M.-S. Alouini is with the Computer, Electrical and Mathematical Science and Engineering Division, King Abdullah University of Science and Technology (KAUST), Thuwal 23955-6900, Saudi Arabia (e-mail: slim.alouini@kaust.edu.sa).}
\thanks{\\ 978-1-5386-5541-2/18/\$31.00~\copyright2022 IEEE. Personal use of this material is permitted. Permission from IEEE must be obtained for all other uses, in any current or future media, including reprinting/republishing this material for advertising or promotional purposes, creating new collective works, for resale or redistribution to servers or lists, or reuse of any copyrighted component of this work in other works.}
%\thanks{Note to Reviewers: Full demo video is also available for reviewers. Please check your manuscriptcentral account. }
	}
	
%\markboth{IEEE Transactions on Vehicular Technology,~Vol.~XX, No.~X, December~2020}{eee}
	
%SHPark, JDPark
	
%	\IEEEoverridecommandlockouts
%\IEEEpubid{\makebox[\columnwidth]{978-1-5386-5541-2/18/\$31.00~\copyright2022 IEEE.  \hfill} \hspace{\columnsep}\makebox[\columnwidth]{ }}
	\maketitle
%	\IEEEpubidadjcol

	\begin{abstract}
Numerous researchers have studied innovations in future sixth-generation (6G) wireless communications. Indeed, a critical issue that has emerged is to contend with society's insatiable demand for high data rates and massive 6G connectivity. Some scholars consider one innovation to be a breakthrough--the application of free-space optical (FSO) communication. Owing to its  exceedingly high carrier frequency/bandwidth and the potential of the unlicensed spectrum domain, FSO communication provides an excellent opportunity to develop ultrafast data links that can be applied in a variety of 6G applications, including heterogeneous networks with enormous  connectivity and wireless backhauls for cellular systems. In this study, we perform video signal transmissions via an FPGA-based FSO communication prototype to investigate the feasibility of an FSO link with a distance of up to 20~km.  We use a channel emulator to reliably model turbulence, scintillation, and power attenuation of the long-range FSO channel. We use the FPGA-based real-time SDR prototype to process the transmitted and received video signals. Our study also presents the channel-generation process of a given long-distance FSO link. To enhance the link quality, we apply spatial selective filtering to suppress the background noise generated by sunlight. To measure the misalignment of the transceiver, we use sampling-based pointing, acquisition, and tracking to compensate for it by improving the signal-to-noise ratio. For the main video signal transmission testbed, we consider various environments by changing the amount of turbulence and wind speed. We demonstrate that the testbed even permits the successful transmission of ultra-high-definition (UHD: 3840 × 2160 resolution) 60 fps videos under severe turbulence and high wind speeds.
    
	\end{abstract}

		\IEEEpeerreviewmaketitle

\begin{figure*}[t]
\centering
{
		\includegraphics[width=1.3\columnwidth]{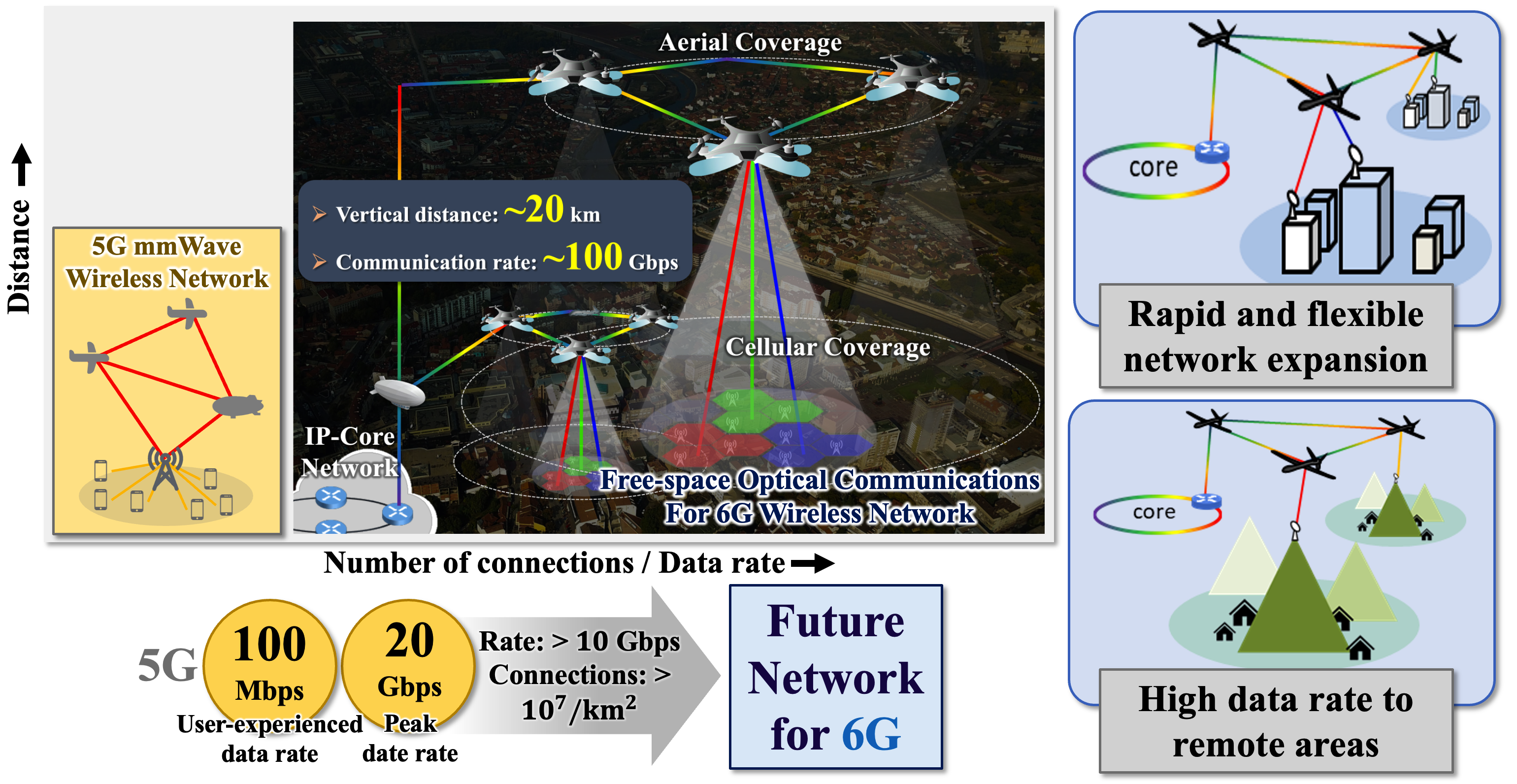}
	}
	\caption{Evolution and use-case scenario of the proposed long-range FSO networks to 6G wireless communications. As the demand for data rate and user connectivity surpasses the capacity of the existing 5G network, researchers have been investigating new spectrum ranges and developing novel transmission strategies to respond to the unprecedented demand.}	\label{6GFSO}
\end{figure*}
\section{Introduction}
Recently, fifth-generation (5G) wireless communication systems have appeared in numerous countries, satisfying the demands of both industries and end users. However, today, researchers have focused on the concepts and applications of future wireless communication, that is sixth-generation (6G) wireless communications. As the roadmap of 6G becomes actualized, the data rates of 6G usage cases, such as streaming services and mobile augmented and virtual reality (AR/VR), are also required at a superior level, which exceeds one terabit per second (Tbit/s). In addition, an increasingly large number of such devices are being connected to the network, and they are communicating with one another. {Such interactions are expected to exceed the connectivity of $10^7$~connections per~$\textrm{km}^2$~\cite{T6G}, thereby surpassing the capacity of 5G (up from $10^6$)~\cite{T6G}.} Moreover, severe problems arise from the scarcity of licensed spectrum resources. {This problem cannot be resolved by existing radio-frequency (RF) communication systems, which are almost saturated by licensed spectrums. Although we find the available RF spectrum and utilize it for mobile access networks, it is difficult to satisfy the target data rate ($\approx$Tbits/s) of 6G because of its relatively low bandwidth~\cite{T6G, UAVFSO}. An emerging solution is the free-space optical (FSO) communication system. FSO communication is a strong candidate for future wireless networks owing to several features, including high capacity and license-free characteristics, that can achieve the targets of 6G by utilizing hundreds of GHz or even THz of bandwidth scales~\cite{QD2}.} For example, accompanied by unmanned aerial vehicles (UAVs), FSO communication can be used as a non-terrestrial wireless backhaul system. Generally, costly wired communication infrastructure (e.g., optical fiber) is necessary to provide reliable network services to rural areas. Many researchers believe that this problem can be solved by deploying FSO-communication-enabled UAVs to provide wireless backhaul connections~\cite{UAVFSO}. Wireless backhauls can be easily developed by combining the mobility of UAVs with the high data rate and massive-connectivity properties of FSO links. Moreover, such backhaul connections can be expanded and reinstalled while adhering to the standards of the 6G network~\cite{T6G, UAVFSO}. %Fig.~\ref{6GFSO} illustrates the evolution and application of FSO communication systems to future wireless networks.

%To ensure the stability of the optical link between ground terminal and the UAV, the pointing, acquisition and tracking (PAT) technique should be considered to guarantee a rapid and sustainable connection. In general, coarse pointing and fine tracking are two main steps in PAT to construct optical links, and those steps are taking charge of acquiring and maintaining the link, respectively. In this paper, first, we suggest our new method of coarse pointing to reduce the chance of link connection outage with the help of radio-frequency (RF) signal, since an optical communication is vulnerable to the atmospheric channel. %We suggest an RF-aided link connection algorithm.
%The idea is to spread RF beacon signals from the air terminal and implement the angle-of-arrival (AoA) estimation algorithm with an RF lens antenna array at the ground terminal~\cite{suk}. The ground station also takes advantage of the global positioning system (GPS) information of the air terminal, which can easily be given through the beacon signal. The estimator adequately utilizes both GPS and estimated angle information. The angle estimation algorithms such as MUltiple SIgnal Classification (MUSIC)~\cite{MUSIC} and EStimation of signal Parameters via Rotational Invariance Techniques (ESPRIT)~\cite{ESPRIT} have deeply been studied; however, lens antenna array-based AoA estimation turned out to be having more potential in terms of accuracy~\cite{CRLB}. 

One of the most significant technical challenges of the FSO communication link is that the channel characteristics--atmospheric turbulence and scintillation--can easily fluctuate the signal by including~\cite{SCI}. To mitigate the effect of this phenomenon, researchers have investigated various methods to construct robust real-time transmission architectures with high bit rates. Most previous studies have developed high-level technologies for high-rate FSO signal transmission. {These technologies have included a 3D video transmission with an adaptive demodulation scheme and a robust channel coding strategy, and a robust UAV-trajectory optimization strategy for UAV-to-ground FSO signal transmission scenario under the turbulent correlated FSO channels~\cite{VTOWC3, JLT2}. They also demonstrated end-to-end transmission of FSO signals with various modulation schemes and link distances of tens or hundreds of meters~\cite{32Q, 11m}.} However, these works solely focused on numerical simulation or short-distance-link prototyping, which cannot cover the actual feasibility of long-range FSO links, widely assumed in various scenarios (e.g., non-terrestrial cellular systems) for 6G wireless networks~\cite{UAVFSO, QD2}.
%and evaluated its performance by transmitting high-definition (HD) video signals.

In this study, we demonstrate a real-time FPGA-based high-resolution video signal transmission via the FSO channel emulator to verify the feasibility of the long-range FSO communication link for 6G. This transmission setup models the channel characteristics of the FSO channel, including turbulence, scintillation, and power attenuation. We processed the video signal via an electrical-to-optical (E/O) converter, optical-to-electrical (O/E) converter, and FPGA module.  The solar background noise critically affects the FSO transceiver and the link quality. To minimize this effect, we applied a spatial selective filtering method that adaptively reduced the field-of-view (FoV) of the receiver and suppressed the background noise generated by sunlight. We applied the proposed sampling-based pointing, acquisition, and tracking (PAT) techniques to improve the signal-to-noise (SNR) ratio to enhance the accuracy of the FSO signal transmission. In the main video transmission testbed, we demonstrate that the proposed communication scheme transmits and processes the video signal, even under harsh channel conditions, such as high turbulence or wind speed. {This study marks the world's first integration of an FSO channel emulator and an FPGA-based module for evaluating an FSO channel's feasibility. As such, our work represents a significant contribution toward the development of 6G mobile networks (e.g., deployment of high-altitude platform (HAP)-assisted backhaul networks) with feasible and long-distance FSO links. We also consider the proposed system as a unified open platform and expect the opportunities for developing advanced link-enhancement technologies (e.g., satellite network in 6G) to be utilized by the proposed platform.}

 {The main structure of this article is as follows: 1)in Section II, we explain why the proposed long-distance FSO link demonstration is needed, including why the vulnerability of the FSO link must be combatted, based on previous studies and their limitations. We end the section by summarizing the mechanism and the contribution of the proposed long-distance FSO link testbed.} 2) In Section III, we present the proposed FSO link prototype, which integrates the FSO channel emulator and the FPGA-based software-defined-radio (SDR) platform and models the long-distance (up to 20~km) FSO link. The proposed testbed includes various techniques to enhance the robustness of the FSO link. These include the space-selective filtering technique, which suppresses the solar background noise, and the sampling-based PAT technique with SNR enhancement. 3) In Section III, we analyze the signal-transmission result using the proposed platform. 4) In Section IV, we present some concluding remarks.

%The rest of the paper is organized as follows. In Section~II, we present the motivations of the emerging FSO communications for the B5G network based on its characteristics and the technical challenges of long-distance FSO links. In Section~III, we propose the system architecture of the FSO communication link prototype, which integrates the FSO channel emulator and the FPGA-based software-defined-radio (SDR) platform and models the long-distance FSO link of up to 20 km. The channel emulator is embedded with the sampling-based PAT technique with SNR enhancement and the space-selective filtering technique, which reduces the solar background noise. These proposed techniques contribute to the significant improvement of the long-distance FSO link quality assumed in the prototype. In Section~IV, the results of the video transmission through our proposed platform are shown. In Section~V, we make some concluding remarks.

%%%%%%%%%%%%%%%%%%%%%%%%%%%%%%%%%%%%%%%%%%
%\begin{figure*}[t]
%	\begin{center}
%		{\includegraphics[width=1.9\columnwidth,keepaspectratio]
%			{Pic/newfigure05.pdf}%
%			\caption{Proposed PAT process and the standard deviation of UAV location estimation.}
%			\label{newfigure}
%		}
%	\end{center}
%\end{figure*}
%\begin{figure}[t]
%	\begin{center}
%		{\includegraphics[width=0.95\columnwidth,keepaspectratio]
%			{Pic/LensStructure.pdf}%
%			\caption{Signal reception by  RF lens antenna arrays.}
%			\label{LensStructure}
%		}
%	\end{center}
%\end{figure}
\begin{figure*}[t]
	\centering
	\includegraphics[width=1.7\columnwidth]{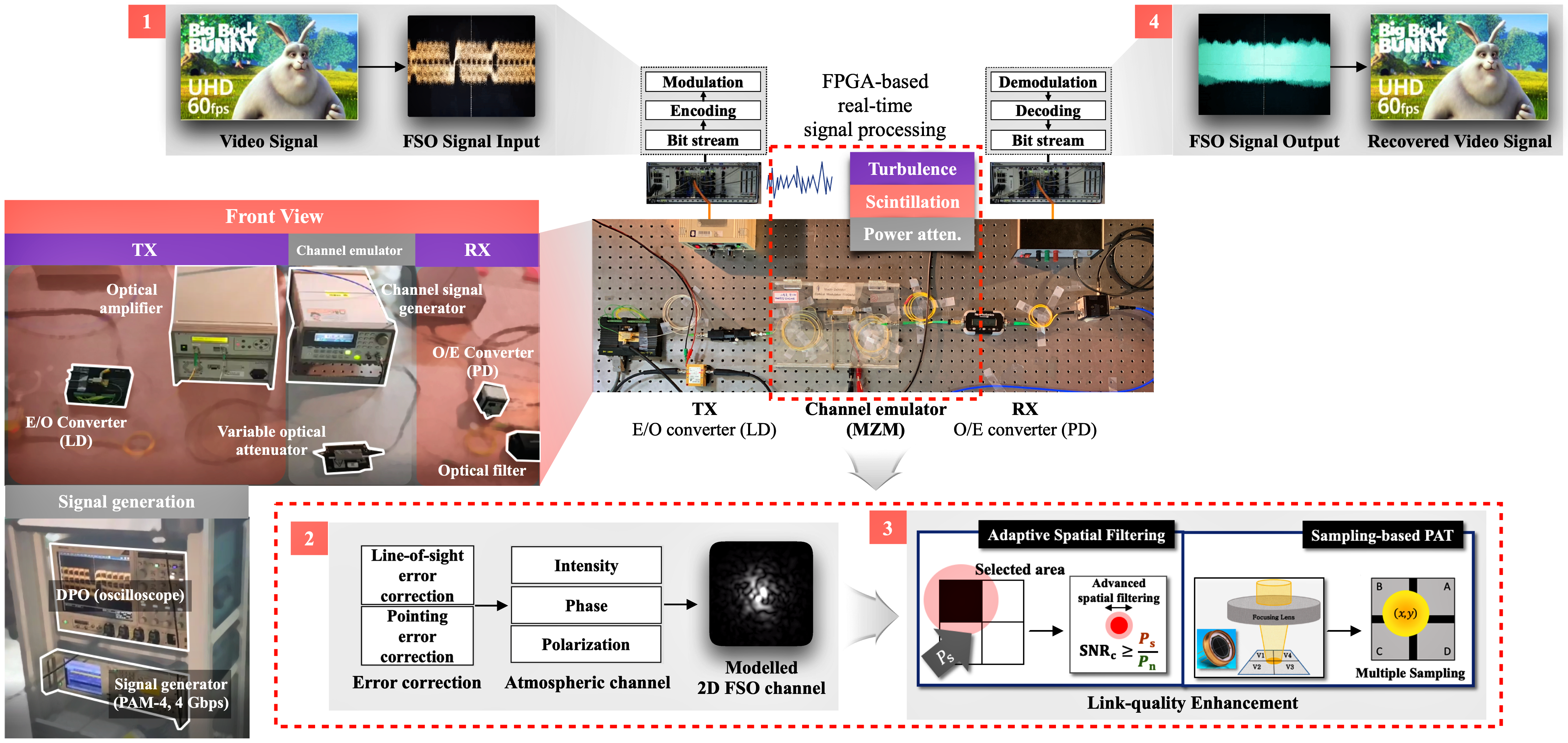}
	\caption{A prototype structure of video signal transmission according to parameter variation related to channel environments and the procedure of proposed feasibility validation of long-distance FSO links.} %Full structure is available in https://bit.ly/37rKXI8.}
	\label{video}
\end{figure*}
\section{Long-Range FSO Communication System for 6G: Challenges and Solution}
{In this section, we explain the confronted challenges to develop the proposed FSO link testbed, including the limitations of previous studies related to the feasibility of a long-range FSO network. Subsequently, we summarize the contribution of the proposed testbed in comparison with previous works.}%review the emergence of FSO communication along with its attendant technical challenges. We also explain the motivation for developing our proposed FSO link testbed.
\subsection{Motivation for the Proposed Real-Time FSO Link Demonstration}

%FSO communication systems use the frequency band of above 300~GHz~\cite{T6G}. Unlike 5G communication system based on RF spectrum, it operates in a near-Terahertz frequency range, which holds a dominant position on obtaining a high data rate for 6G. Indeed, the spectrum over 300~GHz is not licensed worldwide, leading to a considerable saving of the license fees. Moreover, the narrow beam of the FSO signal, which has the order of mrad, can guarantee robustness to interference from other devices and makes it hard to be eavesdropped on, which tightens security~\cite{UAVFSO, SCI}.
\subsubsection{Encountered Challenges of the Long-Distance FSO Link Utilization}
{To utilize the FSO communication which has a dominant position for obtaining a high data rate for 6G wireless networks, as illustrated in the left part of Fig.~\ref{6GFSO}, we must overcome the vulnerability against atmospheric turbulence and other losses, such as penetration, pointing, or propagation losses. These effects become significantly stronger as the distance between the transmitter and receiver increases. For instance, an FSO-communication-based wireless backhaul system with the assistance of UAVs is considered a novel service strategy for 6G, as illustrated on the right side of Fig.~\ref{6GFSO}, assumes a ground-to-UAV link distance of up to 20~km~\cite{UAVFSO}. Therefore, researchers have focused on modeling the turbulence of the FSO signal and developing transceiver techniques to enhance the link quality under various scenarios and reflect this in the link margin strategy~\cite{PAT3, VTOWC3, SCI}.}

\subsubsection{Recent Works and Limitations}
{Owing to its potential in data rate and connectivity, several studies have been conducted to evaluate the actual performance and feasibility of an FSO link. Concurrently, researchers have been striving to develop related technologies for robust FSO communication, including its performance analysis by simulation. In~\cite{VTOWC3}, the authors numerically analyzed the outage performance of FSO fine tracking system that uses multiple passive corner-cube reflectors (CCRs) for spatial diversity and power saving. In~\cite{JLT2}, the authors assumed a UAV-mounted FSO communication system and maximized the flight time of the UAV by optimizing the trajectory based on various atmospheric environments and the channel and rate characteristics of FSO links. Both studies considered the distance of the FSO link up to approximately 1~km. In~\cite{PAT3}, the authors developed an RF lens-antenna-based PAT technique based on an accurate angle-of-arrival (AoA) estimation with a fast steering mirror (FSM) for hybrid RF/FSO communication systems with a maximum link distance of 3~km.}

{To measure the actual feasibility of FSO communications, research continues on how to develop an end-to-end FSO link prototype under various transmission and channel scenarios. The authors in~\cite{32Q} demonstrated an end-to-end link testbed with a distance of 100~m and applied an FSO signal with a 25 GHz data rate and 32-quadrature amplitude modulation (QAM), which is sent  by a 12 wavelength-division-multiplexing (WDM) channel. The authors in~\cite{11m} presented an end-to-end demonstration of transmitting 4- and 64-QAM signals over an FSO channel under non-uniform turbulence, with a maximum distance of 500~m. Recently, the authors in~\cite{RISE} demonstrated a coherent 100 Gbps FSO link with a link distance of only 40 m by applying dual-polarization quadrature phase shift keying (DP-QPSK).}

{As indicated above, existing research works have focused on either overcoming the transmission environments or the link-quality enhancement strategies based on simulation and have assumed a relatively short FSO link of up to a few hundred meters for hardware validation. Nevertheless, formalizing  the challenges and opportunities pertaining to a feasible long-distance FSO link is still considered an open problem. Indeed, it is difficult to measure the exact feasibility or properties of the FSO link for large distances of up to 20~km. This distance also restricts researchers from focusing on analytical expressions and computer simulations.}

\subsection{Contribution of the Proposed Real-Time FSO Link Demonstration}
To solve these problems and evaluate the feasibility of the long-distance FSO link for 6G, we developed a real-time video signal transmission prototype. The prototype is an integration of an FSO channel emulator and an FPGA-based SDR platform. The channel emulator models the time-varying atmospheric channel with power attenuation by absorption and scattering, and it models turbulence with scintillation. An FPGA-based SDR platform was implemented for video signal generation and encoding/decoding. Moreover, we applied various link-quality-enhancement techniques to improve the quality of the received video signal.
 
 {This is the world's first prototype that evaluates the feasibility of the long-distance FSO link by these two essential platforms for FSO and RF communications. Moreover, the novel sunlight noise mitigation and tracking algorithm are jointly included in the testbed for link-quality enhancement, which contributes to the robustness of the FSO signal against the long 20~km transmission distance. Therefore, it can solve the vulnerability issues of long-distance FSO links that occur under atmospheric conditions and transceiver misalignment. By combining these factors, we successfully transmitted an ultra-high-definition (UHD: 3840 $\times$ 2160) video signal over a 20~km FSO link under both clear and hazy atmospheric conditions, owing to the proposed hybrid FSO/RF hybrid platform and outage mitigation techniques. Accordingly, our demonstration significantly contributes to the reliable usage of long-range FSO communication in future wireless networks, including non-terrestrial FSO backhaul networks with a link distance of scale of tens of kilometers, as described in Section~I and Fig.~\ref{6GFSO}.}

\begin{figure*}[t]
	\centering
	\includegraphics[width=1.4\columnwidth]{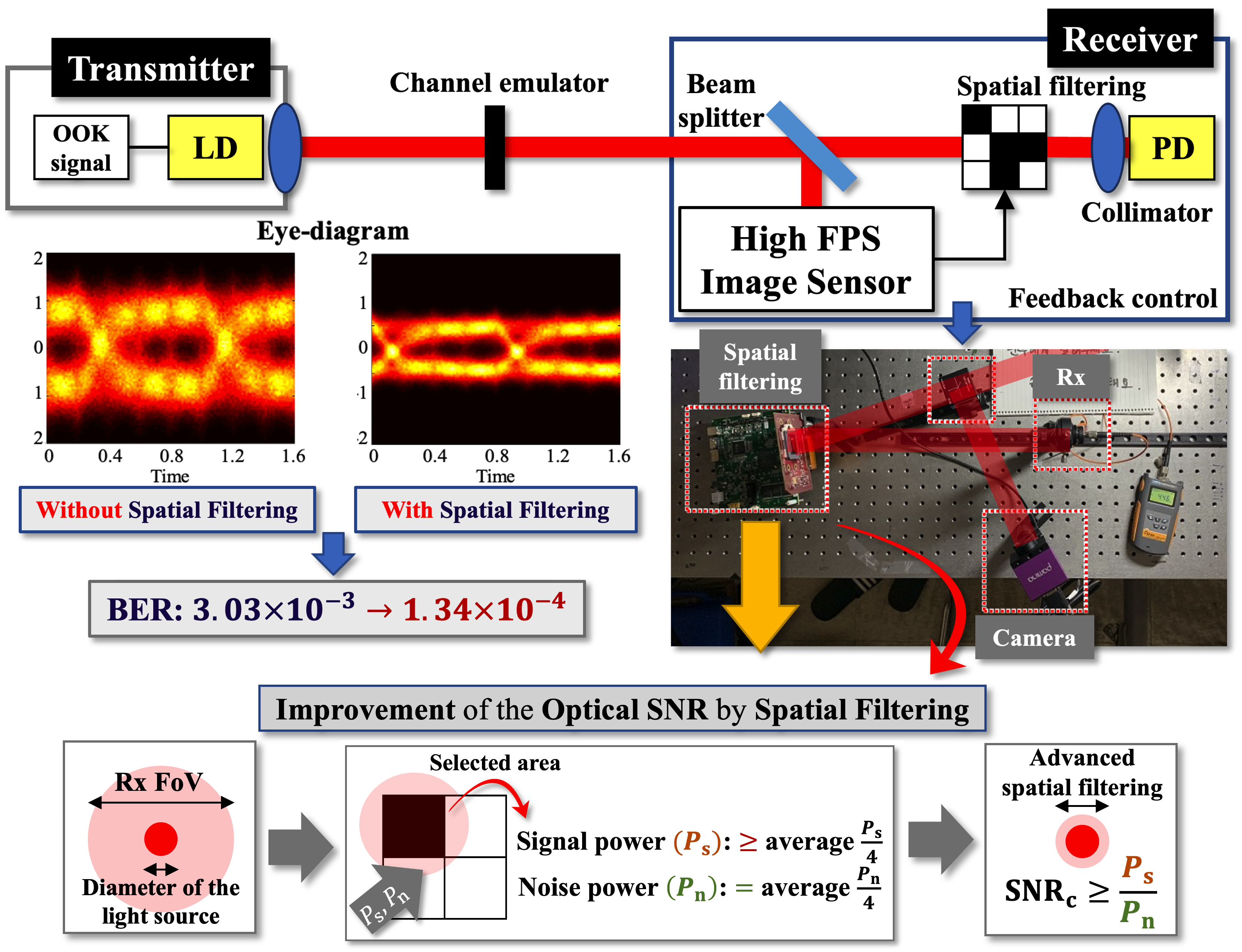}
	\caption{Block diagram and testbed of the proposed spatial selective filtering structure and the results of eye-diagram and BER. Using the proposed adaptive noise suppressing technique, we can significantly lower the noise level and BER.}
	\label{select}
\end{figure*}

\section{System Architecture of FSO Link Prototype}\label{sec3}
In this section, we present the proposed FSO link prototype, including the setup of the FSO channel model, hardware and data, and various techniques for increasing the robustness of up to 20~km in the FSO link. We then analyze the signal transmission results by varying the channel conditions of the experiment.
\subsection{Path-loss Model of the FSO Communication Link}
%We assume both high and low turbulence models when we verify the performance of the estimator. Low and high turbulence channels are supposed to follow log-normal fading and gamma-gamma fading, respectively,~\cite{channel}. 
It is widely known that FSO signals undergo atmospheric attenuation, the degree of which depends on the scintillation and the size of the scattering particles. Pointing inaccuracy takes the center of the beam away from an air terminal, not to receive a maximum intensity, and optical loss occurs owing to the less-than-perfect FSO transceiver elements~\cite{PAT3}. Consequently, the received power is determined by atmospheric attenuation, pointing, and optical loss~\cite{OCS2, UAVFSO}.
\subsubsection{Atmospheric Attenuation: Scintillation}
For atmospheric attenuation, we must consider the scintillation and scattering effects, which are key to the vulnerability of the long-distance FSO link. We employed the scintillation (turbulence) model for atmospheric channel modeling~\cite{SCI, Wind1}. {It represents a function of the refractive index structure parameter $C_n^2$ modeled by the Hufnagel-Valley (H-V) model~\cite{SCI}, which is a function of altitude $h$ in meters and wind speed $v$ in m/s.}
%\begin{equation}
%L_{\textrm{sci}} = 2\sqrt{23.17\left(\frac{2\pi}{\lambda}10^9\right)^{\frac{7}{6}}C_{n}^{2}\ell^{\frac{11}{6}}},
%\label{sci}
%\end{equation}
%where $\lambda$ is the wavelength of the FSO link in nm, $\ell$ is the path length in meters and $C_n^2$ is the refractive index structure parameter modeled by the Hufnagel-Valley (H-V) model, which is a function of the altitude $h$ in meters and the wind speed $v$ in m/s~\cite{SCI}. It is given as follows:
%\begin{equation}
%\begin{split}
%\label{hv}
%C_n^2 = &5.94\times10^{-3} \left(\frac{v}{27}\right)^2 \left(10^{-5} h\right)^{10} e^{- \frac{h}{10^3}} 
%\\&+ 2.7\times 10^{-16} e^{- \frac{h}{1.5\times10^3} } + A_{\mathrm{HV}} e^{- \frac{h}{10^2}}
%\\&\left(A_{\mathrm{HV}}=1.7\times10^{-14} \mathrm{m}^{-\frac{2}{3}}\right).
%\end{split}
%\end{equation}
\subsubsection{Atmospheric Attenuation: Scattering}
We consider the Mie scattering model to express the scattering loss~\cite{Mie2} in dB. This is induced by particles having a size similar to that of the wavelength $\lambda$. As fog, rain, and clouds can cause a scattering effect~\cite{UAVFSO}, a scattering model must be considered for each scenario. For fog and rain, we employed the Kruse model to model the effects~\cite{SCI}. This is a function of the scattering coefficient $\beta_{\textrm{sca}}$ (function of $\lambda$ and visibility range $V$), rainfall rate $R_{\mathrm{r}}$ in mm/h, and the layer thickness of fog and rain $d_{\mathrm{fog}}$ and $d_{\mathrm{rain}}$, respectively. %Here, $\beta_{\textrm{sca}}$ is a .
%It is given by~\cite{SCI}
%\begin{equation}
%\begin{split}
%\label{kruse}
%\begin{cases}
%L_{\textrm{sca}}=4.34\beta_{\textrm{sca}}d~&(\mathrm{fog})\\
%L_{\textrm{sca}}=1.076R_{\mathrm{r}}^{0.67} d~&(\mathrm{rain}),
%\end{cases}
%\end{split}
%\end{equation}
%ince the scattering effect can be occurred by fog, rain and cloud~\cite{channel}, we can consider the ``thickness'' $d$ for each scenario and apply~(\ref{kruse}) to predict the attenuation.
%adopted the Beer-Lambert Law for representing ${L_l}$, which is given by
%\begin{equation}
%\label{attenuation}
%\begin{aligned}
%h_l(z) = \frac{P(z)}{P(0)}=\text{exp}(-\sigma z),
%\end{aligned}
%\end{equation}
%where ${z}$ is a propagation distance, ${P(z)}$ is a power level at distance ${z}$, and ${\sigma}$ is an attenuation coefficient~\cite{visibility}. Here, $\sigma$ can easily be calculated with visibility range $V$~\cite{channel}, which defines a condition of an atmosphere.
In the proposed testbed, ${V =10}$~km and ${V =3}$~km are chosen to represent clear weather and hazy weather, respectively. These values are chosen to verify the transmission results based on two separate Kruse models corresponding to $V\lessgtr6$~km\cite{UAVFSO, Mie2}.
For cloudy conditions, we used the model in~\cite{cloudconf} to express the scattering loss by clouds. Therefore, the total loss in decibels is given by the sum of the path losses of every attenuation factors.
\subsubsection{Pointing and Optical Loss}
We set the pointing and optical losses equal to 2~dB, resulting from the misalignment and optical efficiency of the transceiver, respectively,~\cite{UAVFSO}. The pointing loss is determined by each pointing loss factor of the transceiver, which is determined by the transceiver aperture, wavelength, and pointing error angle~\cite{SCI}. We obtained data in our hardware setup (transceiver aperture and wavelength) and the proposed sampling-based PAT algorithm in Section III.C with low misalignment. The optical loss can be modeled by the decibels of the product of the FSO transmitter and receiver's optical efficiencies $\eta_t$ and $\eta_r$, respectively. The typical value of the product $\eta_t \eta_r$ is in the range of [0.2, 0.7]~\cite{UAVFSO, 32Q}, which depends on the optical components, and a value 0.65 is used in our demonstration, which approximates the optical loss by 2~dB.

\begin{table}[t]
	%\footnotesize
	\centering
	\caption{Parameter setup for video signal transmission through the FSO communication channel}
\begin{tabular}{|c |c|}
			\hline
	%	\toprule
		\textbf{Parameter} & \textbf{Value (hazy, clear weather)}  \\  \hline%\midrule
		%		Scintillation model ($L_{\textrm{sci}}$)                 & As proposed in~\cite{SCI}       \\
		%		Scattering model ($L_{\textrm{sca}}$) & As proposed in~\cite{Mie}\\
		%Pointing and optical loss & 2 dB (equal) \\
		{Wind speed (height: 0~km) $(v)$}        & model in~\cite{Wind1} (6, 1~m/s)       \\\hline
		Visibility $(V)$ & 3, 10~km\\  \hline
		Fog layer thickness $(d_{\mathrm{fog}})$ & 50, 0~m\\  \hline
		Rain layer thickness $(d_{\mathrm{rain}})$ & 1, 0~km\\ \hline
		Cloud attenuation & model in~\cite{cloudconf}\\ \hline
		Wavelength $\left(\lambda\right)$ & 1550~nm\\ \hline
		Link distance $\left(\ell\right)$           & 20~km     \\ \hline
		Bit rate of video signal            & 35~Mbps \\ \hline%200, 48 kb/s \\ 
		%Sample rate            & 44.1~kHz \\
		Resolution and frame rate & UHD 60~fps (H.264)\\ \hline
	\end{tabular}
	\label{param}
\end{table}

\subsection{Hardware and Data Setup}
The prototype structure used to evaluate the feasibility of the FSO link is illustrated in Fig.~\ref{video} and the detailed parameters are listed in Table~\ref{param}. For video signal transmission, we transmitted a one-minute clip of the 2008 animated short \textit{Big Buck Bunny} with 4 K UHD resolution. {Since the transmission is demonstrated in an indoor environment, to induce the accurate signal transmission compared to an actual outdoor scenario with 20~km link distance, we used the Mach-Zehnder Modulator (MZM), which is widely applicable in high-speed optical systems owing to its dominant position on ease of fabrication and flexibility of pulse reputation~\cite{11m, MZM2}. We employed the MZM in the proposed system to model the 2D atmospheric channel and perform error correction, including both line-of-sight (LoS) and pointing errors. Here, using the given atmospheric parameters, we generate the corresponding atmospheric channel with a distance of 20~km by the MZM, which is completed by the following procedure: we divide the transmitted FSO signal into two paths, determine the intensity and phase of the signal by driving voltages to each signal, which causes the variation of optical path length%(e.g., multiplying  $\exp \left(j\frac{\pi V_i}{2V_{\pi}}\right) \left(i=1, 2\right)$ for constant $V_{\pi}$ and symmetric coupler scenario).
and merge them~\cite{MZM2}.} The channel emulator was connected between the transceiver FPGA modules. The transmitter and receiver included a laser diode (LD) with an E/O converter and a photodiode (PD) with an O/E converter, respectively, to ensure long-term video signal transmission. {After receiving an oversampled data stream using a digital sampling oscilloscope (DSO) on the receiver side, we derived the bit-error-rate (BER) performance by using the average and standard deviations of the received ones and zeros~\cite{32Q}. The diodes were connected to an FPGA-based PXIe SDR platform for real-time video signal processing. This involves the encoding and modulation of the bitstream of the video signal. The SDR platform at each transceiver side comprises a PXIe chassis (PXIe-1082) and FPGA controller modules (NI-7976, 5791, PXIe-8880, 8374, and 2953R for the transmitter and receiver sides). %The structure of our FSO link testbed is shown in Fig.~\ref{tb}.
\begin{figure*}[t]
	\centering
	\includegraphics[width=1.98\columnwidth]{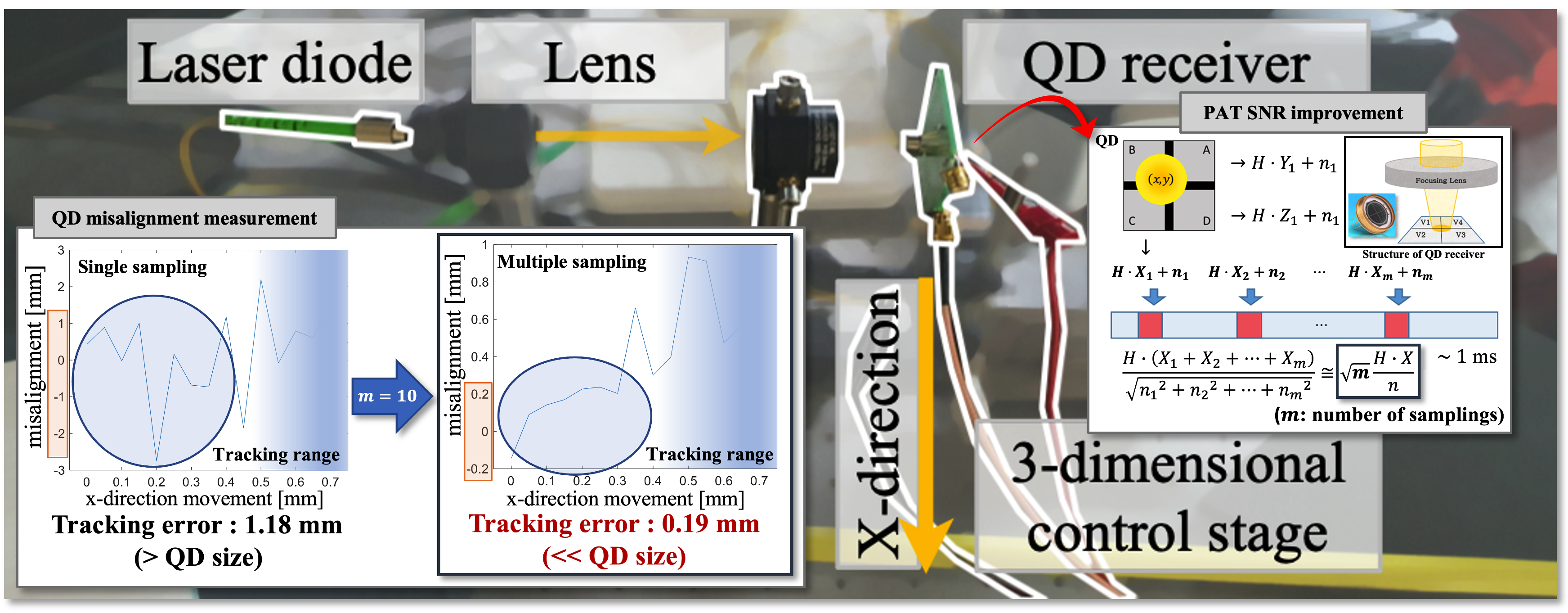}
	\caption{Proposed multiple-sampling-based PAT technique through SNR improving and 3D acquisition and the misalignment result along the x-axis. By enhancing SNR through sampling gain of the square root of the number of samples, we can achieve far less tracking error compared to the size of QD.}
	\label{patx}
\end{figure*}

\subsection{Proposed Techniques for Link-Quality Enhancement}
Since we consider a link distance of up to 20~km, the vulnerability becomes even greater. Hence, we applied several link-quality-enhancing strategies to ensure the reliability of video signal transmission.

\subsubsection{Suppressing Solar Background Noise by Spatial Selective Filtering}
In this section, we present the testbed structure for suppressing the solar background noise. The FSO signal is known to be vulnerable to solar noise~\cite{QD2}, where  % SNR with respect to the received solar radiance is given in Fig.~\ref{srd}. From the figure, we can conclude that indirect sunlight results in
SNR degradation is approximately 5 dB, resulting in more than 50 dB when sunlight becomes direct~\cite{OCS2}. Hence, in the proposed test bed, we suppress this solar background noise compared to the signal power using the proposed spatial selective filtering technique. {We assume that the amount of solar noise is given by (20) in~\cite{QD2}, where the amount of noise is proportional to the receiver FoV, and hence the amount of noise proportionally decays by the selected portion of the receiver aperture.} {Here, we spatially divide the given area into specific partitions (e.g., 2$\times$2, 3$\times$3, $\cdots$) and filter the area where the signal intensity is greater than that of the others. This leads to the result that the signal power of the selected region is, assuming $2\times2$ division, for example, greater than the average signal power $\frac{P_s}{4}$ for the total signal power of aperture $P_s$. However, the noise power of the selected area was suppressed by $\frac{P_n}{4}$ for the total noise power through the total aperture $P_n$. Hence, the optical SNR becomes greater than $\frac{P_{\mathrm{s}}}{P_{\mathrm{n}}}$, which is the SNR of the conventional method without spatial filtering.} To measure the performance of the proposed spatial selection technique, we performed a point-to-point test and realized the algorithm. We split the transmitted FSO link by the beam splitter and determined by our high frames-per-second (fps) image sensor, the amount of reduction of the receiver FoV, and feedback to the receiver. Using the proposed spatial selective filtering method (shown in Fig.~\ref{select}), we can reduce the BER from an order of $10^{-3}$ to $10^{-4}$, which positively contributes to the accuracy of the long-distance video signal~transmission.

\subsubsection{Link-Quality Enhancement by SNR-Improving PAT Technique}
It is widely known that when the FSO transmitter and receiver are misaligned, the link quality is highly degraded, and the outage of the FSO link is increased~\cite{PAT3}. Therefore, we need to apply the proposed multiple-sampling-based PAT technique to our testbed, as illustrated in Fig.~\ref{patx}. To improve the SNR performance, we measured the received $m$ signals for each sampling and acquisition time with the control of the quadrant-detector (QD) receiver, and measured the concatenated optical SNR $\mathrm{SNR_c}$ for given $m$ samples. {Here, the QD receiver detects optical signals from its photoreceiver divided into quarters, estimating the beam displacement by comparing the received signal power of each quadrants~\cite{QD2}. For sampled signals
\begin{equation}
\left\{Y_i\triangleq HX_i+n_i\right\}_{i=1}^m
\label{ss}
\end{equation}
for the static channel, transmitted signal and noise $\left[H, \left\{X_i\right\}_{i=1}^m, n: \left\{n_i\right\}_{i=1}^m\right]$, respectively, the concatenated SNR $\mathrm{SNR_c}$ is given by
\begin{equation}
\mathrm{SNR_c}=\frac{H\cdot\left(X_1+\cdots+X_m\right)}{\sqrt{n_1^2 +\cdots+n_m^2}}\approx \frac{mH\cdot X}{n\sqrt{m}}=\sqrt{\bm{m}} \frac{H\cdot X}{n},
\label{snrc}
\end{equation}
which implies that we can obtain an SNR gain of $\sqrt{m}$ for $m$ sampling compared to a single-signal-reception scenario. Moreover, by our precise control of the QD receiver with angle resolution of sub-$\mu$ rad scale and compensation frequency of up to 1 kHz, the misalignment along the x-axis is reduced by 0.19 mm for ten samplings, which is negligible compared to the size of the QD given by order of~1~mm~\cite{QD2}.

\subsection{Video Signal Transmission under Various Channel Conditions}
%For verification of the channel model's reliability, we conducted an additional demonstration, which is shown in Fig.~\ref{snap} and the results are presented in Table~\ref{rel}. We can see that the bit-error-rate (BER) is $10^{-4}$, which guarantees the reliability of the  2D channel model.
By implementing the proposed testbed using the various link-quality-enhancing techniques described above, we conducted video signal transmission through the emulated FSO channel. {We assumed a four-level pulse-amplitude-modulated (PAM-4) signal for video signal transmission, which achieves superior computational complexity and power consumption for long-distance FSO signal transmission compared to other higher modulation schemes (e.g., $n$-QAM or orthogonal-frequency-division-multiplexing (OFDM)-QAM) that require a complex system architecture and enormous transmit power.} The upper part of Fig.~\ref{snap} illustrates the verification procedure for data reliability by measuring. As shown in the figure, by transmitting a 4-Gbps PAM-4 signal under a weak turbulence channel with a 20-km link distance, our proposed technique can achieve a BER of $10^{-4}$ order. This guarantees the reliability of data transmitted through the 2D channel model.

The lower part of Fig.~\ref{snap} shows snapshots of the video transmission results under the channel conditions. In an environment with low turbulence and slow wind speed, the original clip can be transmitted with almost zero distortion. {Even under harsh conditions of high turbulence and fast wind speed, the clip can be sent with negligible distortion.} Therefore, we can conclude that, with our prototype of a realistic FSO channel model, researchers can confirm the feasibility of the long-distance FSO link under several channel conditions.
\begin{figure*}[t]
	\centering
	\includegraphics[width=1.3\columnwidth]{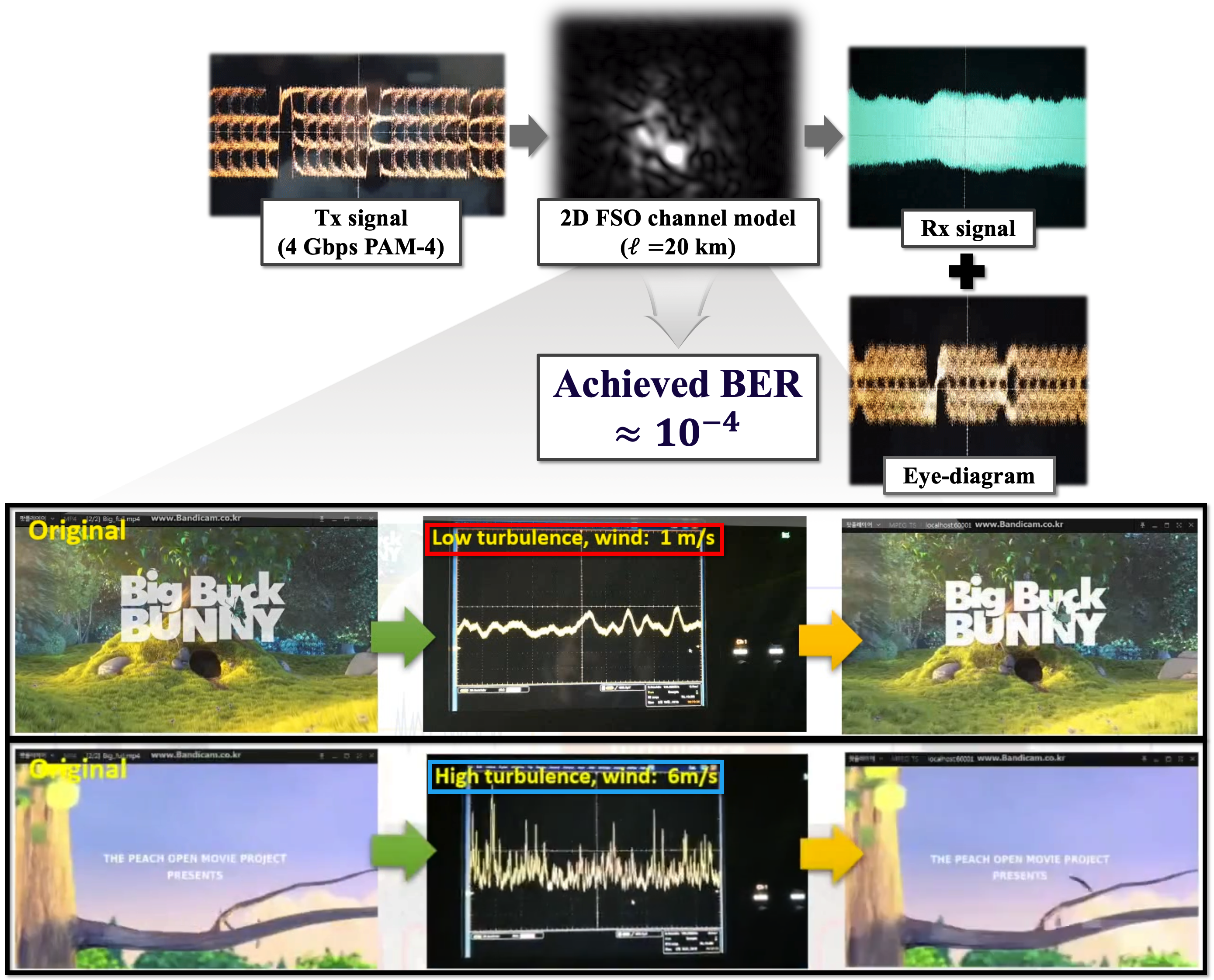}
	\caption{{Data-reliability validation process under $20~\mathrm{km}$ link-distance scenario and snapshots of video transmission in clear (low turbulence, 1~m/s wind speed, ...) and hazy (high turbulence, 6~m/s wind speed, ...) weathers. The denoted wind speed is based on the height of 0~km~\cite{Wind1}.}}
	\label{snap}
\end{figure*}

\section{Conclusion}
In this article, we highlighted the utility of long-range FSO communication links for 6G networks by investigating their high-data-rate, unlicensed, and narrow-beam properties. We also emphasize that the vulnerability of the FSO communication link depends heavily on atmospheric conditions, the harshness of which increases as the link distance increases. To measure the link-level feasibility of the long-distance FSO link, we developed the first novel technique to model a real FSO channel with an MZM emulator and a point-to-point transceiver realized by FPGA modules. We considered the FSO channel with both low- and high-turbulence scenarios and adopted various link-quality enhancement strategies, including solar-noise suppression by selective filtering and SNR-improving PAT. To verify the performance of the strategies, we independently conducted a point-to-point FSO signal transmission experiment and demonstrated how our techniques lowered the BER and misalignment results. For the main feasibility measurement, we demonstrated that even a video signal with UHD 60 fps could be received without distortion (or less) using the proposed SDR-platform-based prototype. We believe that through the description of the challenges and opportunities and the proposed demonstration in this article, we have conveyed a promising insight into the benefits of implementing long-distance FSO links in future wireless networks. Furthermore, we believe that more accurate video transmission techniques, which hold up in practice, can be modeled and tested using the proposed testbed.

%\begin{table}[t]
	%\footnotesize
%	\centering
%	\caption{Parameters for data reliability validation and BER result.}
%	\begin{tabular}{@{}lcc@{}}
%		\toprule
%		Parameter & Information  \\ \midrule
		%		Scintillation model ($L_{\textrm{sci}}$)                 & As proposed in~\cite{SCI}       \\
		%		Scattering model ($L_{\textrm{sca}}$) & As proposed in~\cite{Mie}\\
		%Pointing and optical loss & 2 dB (equal) \\
%		Transmitted signal  & 4~Gbps PAM-4      \\ 
%		Turbulence  & Weak\\
%		Link distance & 20~km\\
%		\textbf{BER} & $\approx\mathbf{10^{-4}}$

%	\end{tabular}
%	\label{rel}
%\end{table}

%---------------------------------------------------------------------------------------------%
%References

\section*{Acknowledgment}
This work was supported by the Institute for Information \& communications Technology Promotion (IITP) grant funded by the Korean government (MSIP) (No. 2019-0-00685, 2022-0-00704).

\bibliographystyle{IEEEtran}
		\bibliography {sample}	

\textbf{Hong-Bae Jeon} is currently pursuing the Ph.D degree in the School of Integrated Technology, Yonsei University, Korea. %His research interests include applied mathematics and emerging technologies for 5G/B5G communications.

\textbf{Soo-Min Kim} is currently pursuing the Ph.D degree in the School of Integrated Technology, Yonsei University, Korea.

\textbf{Hyung-Joo Moon} is currently pursuing the Ph.D degree in the School of Integrated Technology, Yonsei University, Korea.

\textbf{Do-Hoon Kwon} is currently pursuing the Ph.D degree in the School of Electrical and Electronic Engineering, Yonsei University, Korea.

\textbf{Joon-Woo Lee} is currently pursuing the Ph.D degree in the School of Electrical and Electronic Engineering, Yonsei University, Korea.

\textbf{Jong-Moon Chung} is a Professor in the School of Electrical and Electronic Engineering, Yonsei University, Korea.

\textbf{Sang-Kook Han} is a Professor in the School of Electrical and Electronic Engineering, Yonsei University, Korea.

\textbf{Chan-Byoung Chae} is an Underwood Distinguished Professor at Yonsei University, Korea. His research interest includes emerging technologies for 6G and molecular communications.

\textbf{Mohamed-Slim Alouini} is a Distinguished Professor of Electrical Engineering at King Abdullah University of Science and Technology (KAUST), Saudi Arabia. His current research interests include the modeling, design, and performance analysis of wireless communication systems.

\end{document}